\documentclass[iop]{emulateapj}

\usepackage{verbatim}
\newcommand{\kms}{km~s$^{-1}$} 
\newcommand{\water}{H$_2$O} 
\newcommand{\methanol}{CH$_3$OH}

\newcommand{\lsun}{$L_\odot$}
\newcommand{\msun}{$M_\odot$}

\newcommand{\ngci}{NGC~6334I}

\newcommand{\mjb}{mJy~beam$^{-1}$}

\slugcomment{submitted 2017 January 10; revised 2017 January 25; accepted 2017 January 27}

\shorttitle{An Extraordinary Millimeter Outburst in NGC~6334I-MM1}
\shortauthors{Hunter et al.}

\begin{document}

\title{ An extraordinary outburst in the massive protostellar system NGC~6334I-MM1: quadrupling of the millimeter continuum}

\author{
  T. R. Hunter\altaffilmark{1},
  C. L. Brogan\altaffilmark{1}, 
  G. MacLeod\altaffilmark{2}, 
  C. J. Cyganowski\altaffilmark{3}, 
  C. J. Chandler\altaffilmark{4},
  J. O. Chibueze\altaffilmark{5,6,7},
  R. Friesen\altaffilmark{8},
  R. Indebetouw\altaffilmark{1,9},
  C. Thesner\altaffilmark{7},
  K.H. Young\altaffilmark{10}
  }
 
\email{thunter@nrao.edu}

\altaffiltext{1}{NRAO, 520 Edgemont Road, Charlottesville, VA 22903, USA}  
\altaffiltext{2}{Hartebeesthoek Radio Astronomy Observatory, P.O. Box 443, Krugersdorp 1740, South Africa}
\altaffiltext{3}{SUPA, School of Physics and Astronomy, University of St. Andrews, North Haugh, St. Andrews KY16 9SS, UK} 
\altaffiltext{4}{NRAO, P.O. Box O, Socorro, NM 87801, USA} 
\altaffiltext{5}{Department of Physics and Astronomy, Faculty of Physical Sciences, University of Nigeria, Carver Building, 1 University Road, Nsukka, Nigeria}
\altaffiltext{6}{SKA Africa, 3rd Floor, The Park, Park Road, Pinelands, Cape Town, 7405, South Africa}
\altaffiltext{7}{Centre for Space Research, Physics Department, North-West University, Potchefstroom, 2520, South Africa}
\altaffiltext{8}{Dunlap Institute for Astronomy \& Astrophysics, University of Toronto, Toronto, ON, M5S 3H4, Canada}  
\altaffiltext{9}{Department of Astronomy, P.O. Box 3818, University of Virginia, Charlottesville, VA 22903, USA} 
\altaffiltext{10}{Harvard-Smithsonian Center for Astrophysics, 60 Garden Street, Cambridge, MA 02138, USA}


\begin{abstract}

Based on sub-arcsecond Atacama Large Millimeter/submillimeter Array 
(ALMA) and Submillimeter Array (SMA) 1.3~mm continuum images of the
massive protocluster NGC~6334I obtained in 2015 and 2008, we find that
the dust emission from MM1 has increased by a factor of 4.0$\pm$0.3
during the intervening years, and undergone a significant change in
morphology.  The continuum emission from the other cluster members
(MM2, MM4 and the UCHII region MM3=NGC~6334F) has remained constant.
Long term single-dish maser monitoring at HartRAO finds that multiple
maser species toward NGC~6334I flared beginning in early 2015, a few
months before our ALMA observation, and some persist in that state.
New ALMA images obtained in 2016 July-August at 1.1 and 0.87 mm
confirm the changes with respect to SMA 0.87~mm images from 2008, and
indicate that the (sub)millimeter flaring has continued for at least a
year.  The excess continuum emission, centered on the hypercompact HII
region MM1B, is extended and elongated ($1.6'' \times 1.0'' \approx
2100 \times 1300$~au) with multiple peaks, suggestive of general
heating of the surrounding subcomponents of MM1, some of which may
trace clumps in a fragmented disk rather than separate protostars.  In
either case, these remarkable increases in maser and dust emission
provide direct observational evidence of a sudden accretion event in
the growth of a massive protostar yielding a sustained luminosity
surge by a factor of $70\pm20$, analogous to the largest events in
simulations by \citet{Meyer17}.  This target provides an excellent
opportunity to assess the impact of such a rare event on a
protocluster over many years.

\end{abstract}


\section{Introduction}

% SMA data columns extracted from 9:47AM Jun 16, 2016 version of the 2016 ApJ sharelatex.
\begin{deluxetable*}{lcccc}
\tablewidth{0pc}
\tablecaption{ALMA and SMA observing parameters\label{obs}}  
\tablehead{\colhead{Parameter} & \colhead{1.1~mm} & \colhead{0.87~mm} & \colhead{1.3~mm} & \colhead{0.87~mm}\\         &  \colhead{ALMA} & \colhead{ALMA} & \colhead{SMA} & \colhead{SMA}}
\startdata
Observation date(s) & 2016 Jul 18, 2016 Aug 02 & 2016 Jul 17, 2016 Aug 02 &  2008 Aug 17,19 & 2008 Aug 11 \\
Julian day(s) & 2457587.7, 2457603.4 & 2457586.7, 2457603.5 & 2454695.8, 2454700.7 & 2454689.8 \\
Configuration(s) & C36-4, C36-5 & C36-4, C36-5 & Very extended & Very extended \\
Project code & 2015.A.00022.T & 2015.A.00022.T & 2008A-S008 & 2008A-S008\\
Time on Source (min)  & 26, 26 & 27, 27 & 150, 211 & 226 \\
Number of antennas  & 40, 39 & 38, 37 & 8, 6 & 8 \\
FWHP Primary beam ($\arcsec$) & 20 & 17  & 56 & 37 \\
Baseband Freqs. (GHz)  & 280.1, 282.0, 292.1, 294.0 & 337.1, 339.0, 349.1, 351.0 & 220.1, 230.1 & 336.5, 346.5 \\
Spectral windows & 4 & 4 & 48 & 48 \\
Channel width (MHz, km~s$^{-1}$) & 0.9766, 1.0 & 0.9766, 0.85 & 0.8125, 1.1 & 0.8125, 0.71\\
Total bandwidth (GHz) & 7.5 & 7.5 & 3.9 & 3.9  \\
Continuum bandwidth (GHz)  & 0.28 & 0.38 & 1.2 & 0.9 \\
Proj. baseline range (k$\lambda$) & 13-1247 & 14-1424 & 16-380 & 24-588\\
Bandpass calibrator  & J1924-2914, J1517-2422 & J1924-2914 & J1924-2914, 3c454.3 & J1924-2914, 3c454.3\\
Gain calibrator      & J1717-3342 & J1717-3342 & J1733-1304 &  J1733-1304\\
Flux calibrator      & J1733-1304 & J1733-1304 & Callisto  & Callisto \\
Robust clean parameter & $-0.5$ & +0.5 & 0.0 & 0.0 \\  
Resolution ($\arcsec\times\arcsec$ (P.A.$\arcdeg$)) & $0.22\times 0.16$ ($-82$)  & $0.21\times 0.16$ ($-79$) &  $0.84\times 0.33$ ($+16$) & $0.49 \times 0.25$ ($+9$)  \\
RMS noise (\mjb\/)\tablenotemark{a} & 2 & 3 & 7 & 20
\enddata
\tablenotetext{a}{The rms noise varies significantly with position in the images due to dynamic range limitations; the numbers provided here are representative values measured near the sources.}
\end{deluxetable*}

Over the past decade, millimeter interferometers have found many
examples of ``massive protoclusters'', loosely defined as three or
more continuum sources clustered within $\sim$10000~au that appear to
trace massive protostars
\citep[e.g.][]{Cyganowski07,Palau13,Sanchez14}.  The cluster members
typically span a diversity of evolutionary stages ranging from
ultracompact HII~regions and hot molecular cores to cold dust sources.
The cluster member separations are well matched to the Trapezium stars
\citep{Odell09}, suggesting that we are seeing the formation phase of
the central massive stars of future OB clusters.  High resolution
studies of these deeply embedded protoclusters at wavelengths that can
penetrate the high dust column should provide clues to the accretion
mechanisms that ultimately produce massive stars. One such
proto-Trapezium is NGC~6334I, which is located in the northern end of
the Galactic ``mini-starburst'' NGC~6334 \citep{Willis13}.  The
bolometric luminosity of this protocluster is
$\approx1.5\times10^5$~\lsun \citep{Sandell00} using the new maser
parallax distance of 1.30$\pm$0.09~kpc for the neighboring source I(N)
\citep{Chibueze14,Reid14}.  We first resolved NGC~6334I with the
Submillimeter Array (SMA)\footnote{The Submillimeter Array is a joint
  project between the Smithsonian Astrophysical Observatory and the
  Academia Sinica Institute of Astronomy and Astrophysics and is
  funded by the Smithsonian Institution and the Academia Sinica.} at
1.3~mm \citep{Hunter06}, into four sources: the well-known
ultracompact (UC) HII region (MM3=NGC~6334F), two line-rich hot cores
\citep[MM1 and MM2,][]{Beuther07,Zernickel12}, and a dust core MM4.
The latter three objects are so deeply embedded that they remain
undetected at infrared wavelengths at least as long as 18$\mu$m
\citep{DeBuizer02}.  The hot core MM1 lies at the origin of a
high-velocity bipolar molecular outflow from the region
\citep[e.g.][]{Leurini06,Beuther08,Qiu11}, indicating active
accretion.

We observed NGC~6334I with the Atacama Large Millimeter/submillimeter
Array (ALMA) in Cycle~2 at Band~6 (1.3~mm) with an angular resolution
of $0.17''$ (220~au). The continuum from the hot core MM1 was resolved
into seven components within a projected radius of $\approx$1000~au
and with brightness temperatures ranging from 100 to 260~K, consistent
with the gas temperatures implied by the presence of a molecular line
forest and copious \water\/ maser emission \citep{Brogan16}.  The
luminosities inferred from the brightness temperatures and sizes
suggest that MM1 is the dominant member of the protocluster in terms
of bolometric luminosity.  Thus, ALMA's initial view of NGC~6334I MM1
suggested a ``hot multi-core'' of gas and dust containing multiple
massive protostars undergoing accretion.  However, the anomalously
steep 1.5~cm to 1.3~mm spectral energy distributions that we reported
for several components of MM1 motivated further investigation.

In this Letter, we perform a detailed comparison of ALMA images from
2015 and 2016 with SMA images obtained 7-8 years earlier, which
reveals an extraordinary increase in the continuum emission from MM1
while the other protocluster members remained constant. We analyze
several years of single-dish maser spectra and find flaring by a
factor of $>40$ in multiple transitions beginning in early
2015. Finally, using 2016 ALMA target-of-opportunity observations, we
characterize the outburst in terms of luminosity, duration and
location, highlighting its potential impact on the study of episodic
accretion in massive star formation.

\begin{figure*}
\begin{center} 
\includegraphics[width=0.99\linewidth]{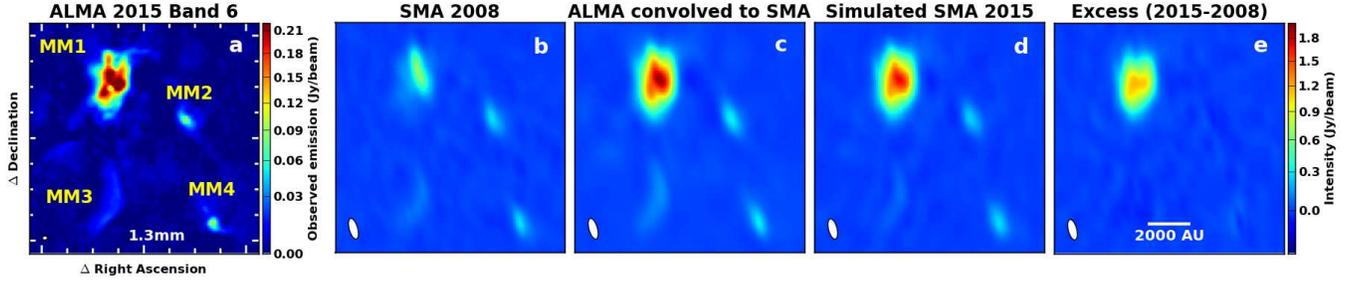}
\caption{\small {\bf (a)} ALMA 1.3~mm continuum observed in 2015 August ($0.20'' \times 0.15''$, Brogan et al. 2016);  {\bf (b)} SMA 1.3~mm continuum observed in 2008 August ($0.84'' \times 0.33''$); {\bf (c)} ALMA image convolved to the SMA beam; {\bf (d)} CASA simulation using the exact $uv$-coverage of panel (b) to observe the sky model of panel (a); {\bf (e)} the simulated SMA image minus the observed SMA image, revealing the large emission excess from MM1 and insignificant change in the other three sources (MM2, MM3, MM4).  Panels (b)-(e) are presented on a common intensity scale. Field center is $17^{\rm h}20^{\rm m}53^{\rm s}.32, -35^\circ47'00''$ (J2000) and field of view is $9''$.\label{fivepanel}}
\end{center}
\end{figure*}

\begin{figure*}
\begin{center} 
\includegraphics[width=0.99\linewidth]{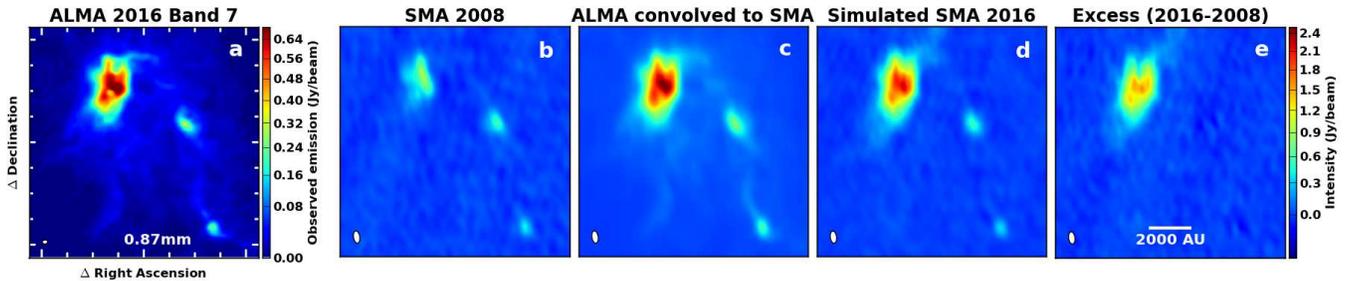}
\caption{\small Same as Fig.~\ref{fivepanel}, but for the 0.87~mm images. Panels (b)-(e) are presented on a common intensity scale.\label{fivepanel2}}
\end{center}
\end{figure*}

\section{Observations}

\subsection{Millimeter and submillimeter interferometry}
\label{obstext}

This Letter includes analysis of new ALMA Cycle~3 observations at 1.1
and 0.87~mm acquired in 2016 via Director's Discretionary Time, along
with older observations using the SMA in 2008 at 1.3 and 0.87~mm. The
observing parameters for these data are presented in
Table~\ref{obs}. The Cycle 3 ALMA data were calibrated using version
C3R4B of the pipeline\footnote{See
  \url{https://almascience.nrao.edu/processing/science-pipeline} for
  details.}. The SMA data were calibrated in MIRIAD and imaged in
CASA.  Because of the strong hot cores in \ngci\/ \citep[MM1 and
  MM2,][]{Zernickel12}, identifying line-free channels suitable for
continuum imaging is challenging. We followed the procedure described
in \citet{Brogan16} and estimate that the resulting images suffer
$<5\%$ residual line contamination toward the hot cores.  The
line-free continuum bandwidths employed are given in Table~\ref{obs}.
The estimated flux scale uncertainties are 5\% for ALMA and 10\% for
SMA.

The continuum data were iteratively self-calibrated, using the bright
emission as the initial model. Multi-scale clean was used on the ALMA
data with scales of 0, 5, and 15 times the pixel size ($0.03''$).
Robust visibility weighting was employed to promote matched angular
resolution in the ALMA data; the beamsizes achieved are given in
Table~\ref{obs}. All images were corrected for primary beam
attenuation.  Our earlier (Cycle~2) ALMA observations of this field at
1.3~mm are also used in this Letter.  They are more fully described in
\citet{Brogan16}, but we include their essential parameters here:
observation date: 2015 August 29 (Julian day 2457264.5), mean
frequency: 230.0 GHz, line-free bandwidth: 0.4~GHz, resolution:
$0.20''\times 0.15''$ (220~AU), rms: 1.2~\mjb.

\subsection{Single-dish maser monitoring}

NGC~6334I is one of several targets in a long-term maser monitoring
program at the Hartebeesthoek Radio Observatory (HartRAO) using the
26~m telescope \citep[see, e.g.,][]{Goedhart04}.  Spectra of the
22.2~GHz \water\/ and 6.7~GHz \methanol\/ maser transitions were
recorded in dual circular polarization with resolutions of 7.81 kHz
(0.11~\kms) and 0.98 kHz (0.044~\kms), respectively, and typical rms
sensitivities of 2~Jy and 1~Jy. The cadence of observations is
typically every 10-15 days.  Flux calibration is performed by
observing Hydra~A and 3C123 assuming the flux scale of \citet{Ott94}.
The telescope beam size is $2'$ and $7'$ at the respective
frequencies.

\section{Results}

\subsection{Millimeter continuum}

We present the ALMA 1.3~mm image from \citet{Brogan16} in
Figure~\ref{fivepanel}a, alongside an SMA image observed seven years
earlier (Fig.~\ref{fivepanel}b).  When we convolve the ALMA image to
the resolution of the SMA image, the result (Fig.~\ref{fivepanel}c)
shows a significant difference in the morphology and the total flux
density of MM1 (10.8$\pm$0.9 vs. 2.34$\pm$0.22 Jy) between the two
epochs.  Interestingly, the other sources in the protocluster (MM2,
MM3, and MM4) do not show significant differences.  As described in
\S~\ref{obstext}, residual line emission contaminating the channels
chosen to make the ALMA continuum image cannot be the cause of such a
large difference in flux density.  Furthermore, the spectrum of the
MM2 hot core is similarly line-rich \citep{Zernickel12} so a line
contamination effect should manifest in that source as well.

To exclude $uv$-sampling effects as the culprit, we simulated the 2008
SMA observations using the 2015 1.3~mm ALMA image
(Fig~\ref{fivepanel}a) as the sky model and applying the identical
$uv$-coverage and flagging of our SMA observations
(Fig~\ref{fivepanel}b) using the CASA {\tt simobserve} task.  To
account for the small difference in frequency between the ALMA and SMA
1.3~mm tunings, we first scaled the sky model image using a spectral
index ($S_{\nu}\propto \nu^{\alpha}$) image constructed from the
contemporaneous 2016 1.1 and 0.87~mm ALMA images.  This technique
corrects for the distinct differences in $\alpha$ among the dust
sources (average values: MM1=2.9, MM2=3.2, MM4=2.4), which would
otherwise cause a systematic error up to 6\%. The resulting image
(Fig.~\ref{fivepanel}d) demonstrates what the SMA would have seen had
it observed the region simultaneously with the 2015 ALMA observations.
We find that the SMA could have recovered $87\%$ (9.4~Jy) of the MM1
emission recovered by ALMA, meaning that the bulk of the difference
between the 2008 and 2015 images cannot be due to
$uv$-sampling. Furthermore, Fig~\ref{fivepanel}e, which shows the
difference between the simulated 2015 SMA image and the actual 2008
SMA image, demonstrates that MM1 exhibits an excess of $f_{\rm
  exc}=6.9\pm0.8$~Jy, while MM2-4 are consistent with being unchanged
($|f_{\rm exc}|< 2.5\sigma$).  We conclude that MM1 has flared by a
factor of 3.9$\pm$0.6 at 1.3~mm during the seven-year interval between the
observations, where the uncertainty includes the combined
uncertainties in the flux scales and image measurements.  This
dramatic increase in flux density explains the anomalously steep
1.5~cm to 1.3~mm spectral energy distributions reported by
\citet{Brogan16} for several of the subcomponents of MM1, and further
indicates that the millimeter flaring happened after the 2011 May VLA
observations.

\begin{figure}[ht!]
%\vspace{-3mm}
\begin{center}
\includegraphics[width=0.99\linewidth]{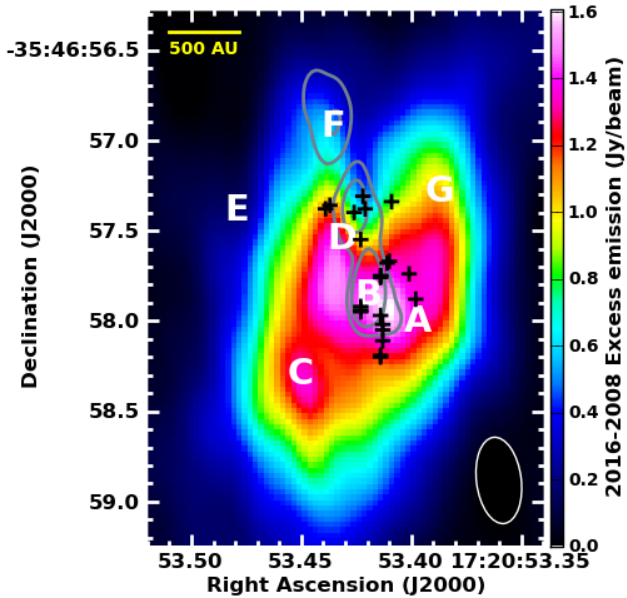}  % built by diff345r0_qband_h2o.py
\caption{\small The 0.87~mm excess emission image of MM1 (2016 compared to 2008; same as Fig~\ref{fivepanel2}e) overlaid with 2011 VLA water maser spots and 7~mm continuum contours (0.24 and 0.6 \mjb; Brogan et al. 2016).   The SMA 0.87~mm beam is shown in the lower right. Letters mark the 1.3~mm continuum components modeled in \citet{Brogan16}.\label{difference} }
\end{center}
%\vspace{-6mm}
\end{figure}

The 2016 0.87~mm ALMA image of NGC~6334I is shown in
Figure~\ref{fivepanel2}a.  The morphology of this image matches well
with the 1.3~mm ALMA image observed one year earlier.  Performing the
same analysis on our 2016 ALMA and 2008 SMA data at 0.87~mm
(Fig.~\ref{fivepanel2}a-e) yields an increase in the flux density of
MM1 from 6.65$\pm$0.68 to 33.3$\pm$1.7 Jy. The SMA could have
recovered $84\%$ (27.8~Jy) of the MM1 emission, yielding a comparable
increase factor of 4.2$\pm$0.5.  Combining the two bands, the mean
factor is 4.0$\pm$0.3; however, the brightening factor could have
evolved over 2015-2016. In any case, in both bands this increase
represents only $\sim$30\% of the total single-dish flux density from
NGC~6334I measured in an $18''$ beam \citep{Sandell94}; thus, it would
not have been so obvious in non-interferometric observations.  In the
highest-resolution image of the excess emission (0.87~mm,
Fig.~\ref{difference}), the hypercompact (HC) HII region MM1B is the
apparent center of activity.  However, the emission is extended and
elongated ($1.6'' \times 1.0'' \approx 2100 \times 1300$~au),
consistent with simultaneous increases in four other subcomponents of
MM1 (A, C, D, and G). The positions of these four subcomponents form an
elongated symmetric pattern relative to MM1B, with a major axis at
position angle +140.5\arcdeg.  This pattern, combined with their
simultaneous brightening, suggests that all of the subcomponents may
not be independent protostars as originally interpreted by
\citet{Brogan16}.

\subsection{Contemporaneous maser flaring}

\begin{figure}[ht!]
\begin{center}
\includegraphics[width=0.99\linewidth]{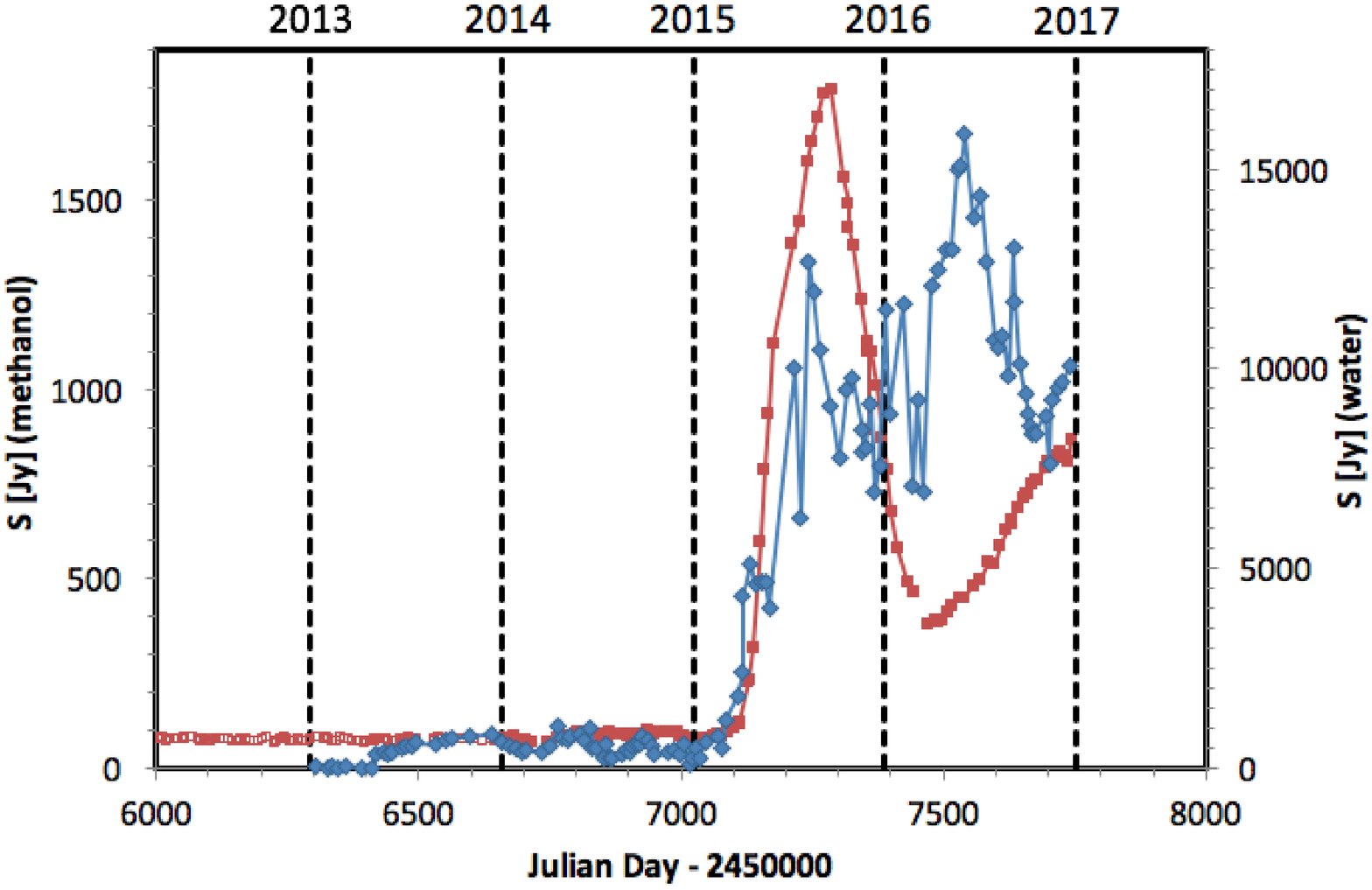}% figure from Gordon, edited in PowerPoint by Todd
\caption{\small Light curves of total intensity from 6.7 GHz methanol (red squares) and 22.2 GHz water (blue diamonds) maser components associated with NGC~6334I. The same LSR velocity component, $-7.2$~\kms, is chosen for each maser species.  The water maser measurements began on 2013 January 26. The methanol masers in NGC~6334~I(N) \citep[located $<2'$ north,][]{Hunter14,Brogan09} are also within the main beam of the 26~m dish at 6.7~GHz, but they appear at a different velocity ($-$2.6~\kms).  \label{masers}}
\end{center}
\end{figure}

Strong flaring in 10 of 15 maser transitions from three species
(\water, \methanol, and OH) toward NGC~6334I began in early 2015
(MacLeod, G. et al. 2017, in preparation).  In Fig.~\ref{masers} we show
light curves for the 22.2~GHz \water\/ and 6.7~GHz Class~II
\methanol\/ maser lines, each for the component at $-7.2$~\kms,
corresponding to the mean LSR velocity of the thermal gas
\citep{Zernickel12}. Both masers had increased by a factor of 10 by
mid-May 2015. The \water\/ maser at this velocity ultimately increased
by a factor of 40 compared to our 2011 May VLA data \citep{Brogan16},
an increase that is reminiscent of past flares in Orion-KL
\citep{Omodaka99,Abraham81}. \water\/ masers in star-forming regions are
believed to occur behind $J$-shocks \citep{Hollenbach13}, suggesting
that an energetic event has occurred. Serendipitously, the 2015 August 
1.3~mm ALMA observations occurred when both masers reached their
initial peak.  While MM2 and MM3 contain 6.7~GHz masers
\citep[e.g.,][]{Brogan16,Krishnan13}, MM1 has never been detected in
this line, nor in the 12.2~GHz Class II maser line, in three decades
of interferometric observations
\citep[][]{Norris88,Norris93,Dodson12}.  Because this maser requires
far-infrared continuum to pump the torsionally excited states
\citep{Sobolev97}, the increase in continuum from MM1 could have
initiated unprecedented maser activity in MM1 and/or enhanced the
existing emission in MM2 and MM3.  Follow-up VLA observations are
underway to pinpoint the flaring masers.

\subsection{Luminosity Increase}

To estimate the increase in luminosity implied by the continuum
flaring of MM1, we first measure the increase in 0.87~mm brightness
temperature ($T_B$) averaged over the region of brightening shown in
Fig.~\ref{difference}.  In an aperture of diameter $1.3''$
($d=1700$~au), $T_B$ has increased by a factor of 2.9$\pm$0.3 (from 33
to 96~K).  Given the uncertain geometry, we consider this change in
the context of a simple model of a central protostar surrounded by a
thick spherical shell of dust.  Since our image resolves this
structure, the dust temperature ($T_d$) will equal
$T_B/(1-e^{-\tau})$, where $\tau$ is the optical depth at 0.87~mm.
The two temperatures should vary proportionally regardless of $\tau$
since the column density and emissivity of the grains (and hence
$\tau$) cannot have changed substantially in such a short period,
except in the hot innermost regions where grains are destroyed (see
\S~\ref{disc}). Thus, we can infer a similar increase factor of 2.9
for $T_d$ at the effective emitting surface.  Radiative transfer
models of dust envelopes surrounding outbursting protostars indicate
that $T_d$ scales as the 1/4 power of the central luminosity
\citep{Taquet16,Johnstone13}; thus, we compute a luminosity increase
factor of $l_{\rm inc}=70\pm20$. We can estimate a lower limit to the
outburst luminosity as the sum of the brightness temperature
luminosity estimates for the five MM1 components affected (A, B, C, D,
and G) reported by \citet{Brogan16}, which is $42000\pm5000$~\lsun.
Applying $l_{\rm inc}$, the pre-outburst luminosity is then
$590\pm170$~\lsun. Assuming the outburst began simultaneously with the
maser burst, its duration is $1.6$~years, so the total energy released
(so far) is $0.8\times10^{46}$~erg.

\section{Discussion}\label{disc}

While unprecedented for a massive star forming core, a rapid
quadrupling of the (sub)millimeter continuum flux density from a
deeply embedded protostellar system like MM1 is within the realm of
theoretical expectation due to episodic accretion, a phenomenon that
is increasingly recognized as being essential to star formation
\citep{Kenyon90,Evans09}. Although dramatic, the $\times$70 luminosity
increase estimated for MM1 is only about double that of the first
identified outbursting Class~0 low-mass protostar HOPS~383 in Orion
\citep[$\times$30-50,][]{Safron15}, whose submillimeter flux density
increased by a factor $>2$.  Based on continuum imaging alone, the MM1
increase occurred sometime during the 4.3~year interval between the
VLA observations and the ALMA 1.3~mm observations.  Since the
initiation of the maser flare lies within this window, it likely
further constrains the rise time.  If the maser and continuum flares
began simultaneously, then the rise time was less than a few months,
compared to the $\approx$18~month constraint for HOPS~383 from its
mid-infrared measurements.

\citet{Johnstone13} simulate the effect of accretion outbursts on the
envelope surrounding a low-mass protostar. A $\times$100 increase in
luminosity yields approximately a $\times$5 increase in the 0.87~mm
flux density after 208~days, which is somewhat faster and larger than
the factor of 4.2 seen in MM1 at 445~days after the maser flare.  The
discrepancy may be due to longer timescales in a higher mass system
where the dust will be heated to much higher temperatures in the inner
optically thick zone.  Indeed, the dust in the innermost region will
be evaporated as the radius of the dust sublimation temperature
\citep[$\sim$1500~K; e.g.,][]{Vaidya09} moves outward, thereby
enriching the gas phase \citep{Taquet16}.  The timescale to transfer
heat from the surviving dust to the gas will be longer than the
initial dust heating time \citep{Johnstone13}.  Modeling the rich,
severely blended line emission from complex organic species in order
to measure the ongoing changes in gas temperature and chemical
abundance compared to the pre-burst state will be a challenging but
important avenue of study.

The primary mechanism that could account for such a large outburst is
a rapid increase in the accretion rate due either to disk
fragmentation \citep[][]{Dunham12,Tobin16} or an encounter
\citep{Pfalzner08a}.  A more extreme possibility is a protostellar
merger, though for massive protostars this phenomenon would typically
produce a much larger energy release than observed \citep{Bally05}.
In the theory of competitive accretion \citep{Bonnell05}, multiple
seeds accrete gas simultaneously with higher rates occurring at the
center of the protocluster. Simulations of such clustered environments
find that encounter-induced boosting of the disk accretion rate is
possible \citep{Pfalzner08}, including up to $\times 1000$
\citep{Forgan10}.  While there are several massive protostars with
candidate disks \citep[see the list in][]{Forgan16}, the first
evidence for a disk-mediated accretion burst from a high-mass young
stellar object was recently reported using infrared observations
toward S255IR~NIRS3 \citep{Caratti17}.  It is similarly massive but
less deeply embedded than MM1, enabling better constrained luminosity
measurements.  While its luminosity increase factor ($\times$5.5) is
lower than MM1, its initial luminosity ($2.9\times10^4$~\lsun) is
significantly larger.  Nevertheless, the two events are quantitatively
similar in terms of the rapid onset, sustained duration
($\gtrsim1$~yr), and total energy, which points toward a common
origin.  Both objects appear to drive bipolar outflows, which
nominally require central disks.  Intriguingly, the position angle of
the elongated MM1 excess emission is exactly perpendicular to the
large-scale bipolar outflow \citep[+50\arcdeg,][]{Brogan16} and the
quasi-symmetry of the excess with respect to MM1B suggests a nearly
edge-on, thick disk-like structure.  Thus, MM1A, C, D, and G may trace
clumps in a fragmented disk heated by an accretion event onto MM1B,
which would explain their co-brightening.  However, in this scenario,
whether any of these components also contain central protostars is
uncertain from the present data, since their high $T_B$ could be
consistent with external heating by a $\sim10^5$~\lsun\/ central
source after accounting for a modest beaming factor ($\gtrsim$2) from
the ``flashlight effect'' \citep{Yorke99}, an anisotropic radiation
field favoring the poles that develops in simulations of massive
accretion disks \citep[e.g.][]{Kuiper15,Klassen16}.

The large, sustained increase in luminosity is analogous to FU Ori
outbursts \citep{Hartmann96}, whose decay periods range from 10 to
several 100~years \citep{Vittone05,Audard14}, during which time an
elevated quasi-steady accretion rate can persist
\citep[e.g.,][]{Zhu10}. It is also consistent with HOPS~383, which
shows no fading after 6~years \citep{Safron15}. In recent numerical
hydrodynamic simulations of massive protostellar accretion including
gas self-gravity and radiative feedback \citep{Meyer17}, the central
protostar undergoes a burst in accretion rate from 10$^{-3}$ to
10$^{-1}$~\msun~yr$^{-1}$ as a massive (0.55~\msun) gas fragment
approaches and enters the protostellar sink cell.  Spaced by
$\sim$2000~years, these largest events generate luminosity bursts of
$\times$50-100 lasting for several years, two characteristics
analogous to the MM1 event.  To test this scenario, it will be crucial
to monitor the evolution of MM1 in future years in both dust and
free-free emission, as well as to discern if there has been any change
to the outflow properties due to the outburst.  Toward this end, we
have higher-resolution ALMA observations in the Cycle 4 queue, which
should also help elucidate the nature of the MM1 subcomponents.

\acknowledgments

The National Radio Astronomy Observatory is a facility of the National
Science Foundation operated under agreement by the Associated
Universities, Inc. The Dunlap Institute is funded through an endowment
established by the David Dunlap family and the University of Toronto.
This Letter makes use of the following ALMA data:
ADS/JAO.ALMA\#2013.1.00600.S, 2015.A.00022.T. ALMA is a partnership of
ESO (representing its member states), NSF (USA) and NINS (Japan),
together with NRC (Canada) and NSC and ASIAA (Taiwan) and KASI
(Republic of Korea), in cooperation with the Republic of Chile. The
Joint ALMA Observatory is operated by ESO, AUI/NRAO and NAOJ.  CJC
acknowledges support from the STFC (grant number ST/M001296/1).  This
research used NASA's Astrophysics Data System Bibliographic Services.

Facilities: \facility{ALMA}, \facility{SMA}, \facility{HartRAO}.

\clearpage

%% the file, uncomment the \plotone, \plottwo, and
%% If you need a layout that cannot be achieved with \plotone or
%% \plottwo, you can invoke the graphicx package directly with the
%% \includegraphics command or use \plotfiddle. For more information,
%% please see the tutorial on "Using Electronic Art with AASTeX" in the
%% documentation section at the AASTeX Web site,
%% http://www.journals.uchicago.edu/AAS/AASTeX.
%%
%% The examples below also include sample markup for submission of
%% supplemental electronic materials. As always, be sure to check
%% the instructions to authors for the journal you are submitting to
%% for specific submissions guidelines as they vary from

%% This example uses \plotone to include an EPS file scaled to
%% 80% of its natural size with \epsscale. Its caption
%% has been written to indicate that additional figure parts will be
%% available in the electronic journal.
\end{document}